# MESSENGER observations of planetary ion enhancements at Mercury's northern magnetospheric cusp during Flux Transfer Event Showers

Weijie Sun[1*], James A. Slavin[1], Anna Milillo[2], Ryan M. Dewey[1], Stefano Orsini[2], Xianzhe Jia[1], Jim M. Raines[1], Stefano Livi[1,3], Jamie M. Jasinski[4], Suiyan Fu[5], Jiutong Zhao[5], Qiu-Gang Zong[5], Yoshifumi Saito[6], Changkun Li[1]

[1] Department of Climate and Space Sciences and Engineering, University of Michigan, Ann Arbor, MI 48109, USA

[2] Institute of Space Astrophysics and Planetology, INAF, via del Fosso del Cavaliere 100, 00133, Rome, Italy

[3] Southwest Research Institute, San Antonio, TX, USA

[4] NASA Jet Propulsion Laboratory, California Institute of Technology, Pasadena, CA, USA

[5] School of Earth and Space Sciences, Peking University, Beijing, China

[6] Japan Aerospace Exploration Agency, Institute of Space and Astronautical Science, Kanagawa, Japan

*corresponding to Weijie Sun (wjsun@umich.edu)

## Abstract

At Mercury, several processes can release ions and neutrals out of the planet's surface. Here we present enhancements of dayside planetary ions in the solar wind entry layer during flux transfer event (FTE) "showers" near Mercury's northern magnetospheric cusp. The FTE showers correspond to the intervals of intense magnetopause reconnection of Mercury's magnetosphere, which form a solar wind entry layer equatorward of the magnetospheric cusps. In this entry layer, solar wind ions are accelerated and move downward (i.e. planetward) toward the cusps, which sputter upward-moving planetary ions within 1 minute. The precipitation rate is enhanced by an order of magnitude during FTE showers and the neutral density of the exosphere can vary by >10% due to this FTE-driven sputtering. These *in situ* observations of enhanced planetary ions in the entry layer likely correspond to an escape channel of Mercury's planetary ions, and the large-scale variations of the exosphere observed on minute-timescales by Earth observatories. Comprehensive, future multi-point measurements made by BepiColombo will greatly enhance our understanding of the processes contributing to Mercury's dynamic exosphere and magnetosphere.

**Key Points.**

1. Solar wind entry layer is formed equatorward of cusp during FTE showers, which enhances precipitation rate by an order of magnitude
2. Solar wind sputtering can release planetary ions and neutrals efficiently within 1 minute of the onset of magnetopause reconnection
3. A new escape channel of planetary ions is formed by solar wind sputtering with an escape rate of $10^{23}$ $s^{-1}$ for $Na^+$ ions

## 1. Introduction

Mercury possesses a global dipole magnetic field with a similar polarity of Earth's dipole field, but the magnetic field intensity near Mercury's magnetic equatorial plane (~ 200 nT) is much less than the strength of Earth's field (~ 30,000 nT) [see, *Anderson et al.*, 2012]. Mercury's magnetic field can hold off the constantly streaming solar wind with a subsolar magnetopause distance of around one thousand kilometers above the planet's surface [*Siscoe et al.*, 1975; *Slavin et al.*, 2008]. As the closest planet to the Sun, Mercury is subject to the strongest solar wind driving compared to other planets in the solar wind [*Slavin & Holzer*, 1981; *Sun et al.*, 2022]. One outcome is that the magnetic reconnection erosion effect on the dayside magnetopause is significant at Mercury [*Slavin & Holzer*, 1979; *Slavin et al.*, 2014; *Slavin et al.*, 2019], and it often generate flux transfer events (FTEs) on the magnetopause [*Russell & Elphic*, 1978; *Russell & Walker*, 1985]. In Mercury's magnetosphere, the FTEs are observed often in quick succession and a large number by MErcury Surface, Space ENvironment, GEochemistry, and Ranging (MESSENGER [*Solomon et al.*, 2007]), which is named flux transfer event (FTE) "shower" [*Slavin et al.*, 2012; *Sun et al.*, 2020b].

The occurrence of FTE showers is high and depends on both of the magnetosheath plasma plasma $\beta$, which is the ratio of the thermal pressure to the magnetic pressure, and the magnetic shear angle across the magnetopause [*Sun et al.*, 2020b]. The high occurrence of FTE showers at Mercury is due to Mercury's magnetosphere being embedded in the solar wind with a low Alfvénic Mach number ($M_A \lesssim 4$) [*Slavin & Holzer*, 1981; *Sun et al.*, 2022], i.e., the ratio of solar wind speed to the Alfvén speed. The low $M_A$ solar wind often leads to a magnetosheath with a low plasma $\beta$ that forms a plasma depletion layer (PDL) ahead of the magnetopause [*Gershman et al.*, 2013; *Slavin et al.*, 2014]. As a result, magnetic reconnection at Mercury's magnetopause would be less dependent on the polarity of the IMF [*Swisdak et al.*, 2010] and occur in high reconnecting rates [*DiBraccio et al.*, 2013]. This can be revealed by the fact that not only do FTE showers occur in high occurrence rates (around 52% of the magnetopause crossings) but also the FTEs are observed in quick succession, i.e., separated by only a few seconds. FTEs occur at lower occurrence rates and are separated by tens of minutes at other planets [*Russell*, 1995; *Jasinski et al.*, 2021a], such as Earth, Jupiter, and Saturn since they are embedded in the solar wind with a higher $M_A$.

Magnetic field lines in the FTEs are open field lines with one end connecting to the solar wind and the other to the planetary magnetic field to the magnetospheric cusp [*Lee & Fu*, 1985]. The solar wind particles can be accelerated and transported into the magnetosphere and bombard the regolith beneath the cusps, which correspond to the process of solar wind sputtering [*McGrath et* *al.*, 1986; *Hofer*, 1991; *Killen et al.*, 2001; *Mura et al.*, 2005]. Mercury does not have a significant atmosphere but a tenuous exosphere. Since the discovery of Mercury's neutral exosphere [*Broadfoot et al.*, 1974; *Potter & Morgan*, 1985], thermal and photon-stimulated desorption [*Sprague et al.*, 1997; *Madey et al.*, 1998], micrometeoroid impact vaporization [*Morgan et al.*, 1988; *Mangano et al.*, 2007; *Jasinski et al.*, 2020] and the solar wind sputtering have been proposed to release neutral particles [*McGrath et al.*, 1986; *Hofer*, 1991; *Killen et al.*, 2001; *Mura et al.*, 2005] and ions [*Broadfoot et al.*, 1974; *Benninghoven*, 1975; *Raines et al.*, 2014] from the planet's surface. However, these processes and their relative importance are poorly understood.

Here we present MESSENGER's observations of the enhancement of planetary ions, specifically of the $Na^+$-group ions, which includes $Na^+$ (sodium ion), $Mg^+$ (magnesium ion), $Al^+$ (aluminum ion) and $Si^+$ (silicon ion), near Mercury's northern magnetospheric cusp during flux transfer event (FTE) "showers". The FTEs accelerate and magnetically channel solar wind protons downwards and planetwards toward the magnetospheric cusps, which forms a solar wind entry layer. This entry layer is observed to increase the proton precipitation rate beneath the cusps at the planet's surface by an order of magnitude. This produces anti-planetward-moving planetary ions within around 1 minute after the onset of an FTE shower. The neutral density of the exosphere can vary by >10% due to this FTE-driven sputtering. These *in situ* observations of enhanced planetary ions in the entry layer likely correspond to an escape channel of Mercury's planetary ions, and the large-scale variations of the exosphere observed on minute-timescales by ground-based solar telescopes.

## 2. Satellite and Instrumentation

MESSENGER orbited Mercury between March 18, 2011, and April 30, 2015 UTC, around seventeen Mercury years. In this study, measurements of the magnetic field made by the Magnetometer (MAG) [*Anderson et al.*, 2007] and of ions by the Fast Imaging Plasma Spectrometer (FIPS), which is a part of Energetic Particle and Plasma Spectrometer (EPPS) [*Andrews et al.*, 2007], were used. The MAG provides the magnetic field at a time resolution of 20 vectors per second in the Mercury solar orbital (MSO) coordinate system.

The Fast Imaging Plasma Spectrometer (FIPS) was an ion-mass spectrometer, which could resolve mass per charge from 1 to 60 amu/e through energy per charge (E/q) and time-of-flight (TOF) measurement. The range of energy per charge of FIPS was from about 0.1 to 13.5 keV/e with a time resolution of ~ 10 s (scan time is ~ 8 s). FIPS used the double-coincidence technique, which greatly reduced background noise. FIPS had a field-of-view (FOV) of around 1.4$\pi$ sr and an angular resolution of 10°. However, the spacecraft obstructions reduce the FOV to an effective value of ~ 1.15$\pi$ sr.

Sodium group ion data used in this study contain ions of mass per charge from 21 to 30 amu/e, which include $Na^+$, $Mg^+$, $Al^+$, and $Si^+$ [*Raines et al.*, 2013]. The measurements were binned to increase the signal-to-noise ratio, and it is not possible to directly distinguish between these species at present. The $Na^+$-group ions in this paper refer to the above-mentioned group of species. The densities of $Na^+$-group ions and the Alpha particle ($He^{++}$) used in this study are the observed number density. The observed density is calculated by integrating the phase space density, which is converted from the observed counts, over the observed velocity range and FOV. The real density should be about 3.48 (4$\pi$/1.15$\pi$) times the observed density. A detailed description of the FIPS, including its FOV, double coincidence technique, and derivation of observed density, was given by [*Raines et al.*, 2011; *Raines et al.*, 2013].

The aberrated Mercury solar orbital (aMSO) coordinates were used for the spacecraft position. In the MSO coordinates, the $\hat{x}_{MSO}$ points from the center of Mercury to the Sun, the $\hat{z}_{MSO}$ points northward perpendicular to the Mercury's orbital plane, and the $\hat{y}_{MSO}$ completes the right-handed system, which is directed opposite to Mercury's orbital velocity around the Sun. In the aMSO, the coordinates of $\hat{x}_{aMSO}$-$\hat{y}_{aMSO}$ plane is rotated to be antiparallel to the solar wind (400 km/s along -$\hat{x}_{aMSO}$).

The aberrated Mercury solar magnetospheric (aMSM) coordinates were used for the magnetic field data. Because the dipole axis of Mercury is close to the $\hat{z}_{MSO}$ with an angle difference of < 0.8°, the three-axis of aMSM are the same as those of aMSO except that the origin of aMSM is at the center of the dipole, which is shifted northward of approximately 484 km from the planetocentric position.

## 3. FTE shower on 22 March 2012

### 3.1. Event Overview

On the left side of Figure 1, the measurements of the magnetic field and ions during one of the MESSENGER's descending orbits on 22 March 2012 are displayed. In around 20 minutes, MESSENGER travelled from the subsolar magnetopause to the northern cusp and then reached the closest approach (CA) at an altitude of around 260 km, during which MESSENGER observed abundant $Na^+$-group ions (including $Na^+$, $Mg^+$, $Al^+$, and $Si^+$) with the observed densities ($n_{obs}$) of 0.5 $cm^{-3}$ to 2 $cm^{-3}$ (Figure 1c). Early in the time series, near the magnetopause,

MESSENGER observed approximately 40 FTEs in less than five minutes. These FTEs, separated by only a few seconds, appear in a large group, which correspond to an interval of FTE shower.

During this event, the interplanetary magnetic field (IMF) in the magnetosheath was southward with a magnetic shear angle across the magnetopause of 107°. The magnetic field intensity in the magnetosheath was slightly smaller than the field intensity in the magnetosphere, which implied a plasma depletion layer (PDL) ahead of the magnetopause [*Gershman et al.*, 2013], and the plasma $\beta$ was calculated to be approximately only 0.08 in the magnetosheath. As the plasma $\beta$ in the dayside magnetosphere was small (<0.1), the magnetic reconnection was approximately symmetric at this magnetopause.

This event at Mercury's magnetosphere was also modeled by a global Hall-Magnetohydrodynamics (MHD) simulation with a coupled planetary interior [*Jia et al.*, 2015; *Chen et al.*, 2019; *Jia et al.*, 2019] under similar low magnetosheath plasma $\beta$ and southward IMF conditions. See Appendix A on the introduction of the Hall-MHD simulation. On the right hand side of Figure 1, the magnetic flux ropes centered in the FTEs with helical magnetic field lines are gathered on the dayside magnetopause. These flux ropes are formed between neighboring X-lines, in which the magnetic field lines have one end in the solar wind while the other passes through the northern or southern cusps and down into Mercury's surface. Consequently, for the present case in the northern hemisphere, the solar wind particles parallel to the magnetic field lines would travel into the magnetosphere along the open magnetic field lines inside the FTEs. In Figure 1a, the intermittently appearing cold ($\lesssim$ 1keV) and dense protons in the high latitude magnetosphere correspond to these injected ions. These injected solar wind protons are further analyzed in the next section. In the lower altitude magnetosphere, including the cusp, the injected solar wind particles along the open field lines inside the FTEs diamagnetically reduce the planetary magnetic field and generate magnetic depressions, which are known as filaments (see Figure 1g and ref. [*Slavin et al.*, 2014; *Poh et al.*, 2016]).

### 3.2. Injected Solar Wind Protons

Figure 2a shows the proton phase-space-density (PSD) distribution versus pitch angles, which integrates the FIPS measurements inside the FTEs in Figure 1. FIPS provided the proton distribution in pitch angles from 30° to 150° (with more than 40 scans). The cold and dense protons that appeared in the parallel direction (pitch angles smaller than 90°) corresponded to the injected magnetosheath protons along the open magnetic field lines in the FTEs. As shown in Figure 2b, the cold and dense protons with pitch angles from 40° to 60° (the red line) corresponded to a temperature of around 0.66 keV and a density of around 22.2 $cm^{-3}$, which were consistent with the features of magnetosheath protons. Considering the characteristic energy of 0.66 keV, these high flux solar wind protons would take only around 10 seconds to travel from the subsolar point to the planet's surface beneath the cusp. These high flux solar wind protons would eventually bombard the planet's surface and cause sputtering.

Figure 3 shows the particle flux versus pitch angles of the $Na^+$-group ions during the period from 03:57 to 04:08, 22 March 2012 UTC. This figure integrated the measurements of the $Na^+$-group ions in the dayside magnetosphere. Figure 3a shows that more of the $Na^+$-group ions were moving anti-planetward than planetward, which suggests that the planetary ions were generated and outflowing from the planetary surface. We show and discuss the outflowing $Na^+$-group ions in the next section about statistical analysis.

## 4. Statistical Analysis

### 4.1. Spatial Distribution of $Na^+$-group Ions and Magnetic Field Line

In the entire mission, MESSENGER traversed the dayside magnetopause around 3748 times, in which 1953 (about 52%) were accompanied by FTE showers (≥ ten FTEs during the magnetopause crossing) [*Sun et al.*, 2020b]. Intervals when MESSENGER crossed the dayside magnetosphere (normally around 15 minutes) where FTE showers were observed were defined as FTE shower intervals (1953 events). The remaining dayside magnetosphere crossings (1795 events) without FTE showers were defined as non-FTE shower intervals. It is hard to know the magnetopause reconnection conditions in real time with only MESSENGR observations. A study focusing on IMF near Mercury's orbit shows that the IMF is likely to retain a similar state for 10 to 20 minutes [*James et al.*, 2017]. Therefore, using the FTEs as an indicator of magnetopause reconnection is appropriate for our study.

Figure 4 shows the spatial distributions of the observed density ($n_{obs}$) of the $Na^+$-group ions near the noon-midnight meridian plane ($R_{xy}$-$Z$) plane during the non-FTE shower intervals (Figure 4a) and the FTE shower intervals (Figure 4b), which are overlaid with the magnetic field lines (in white) during their intervals, respectively. Appendix B described on how the magnetic field lines were derived. The $Na^+$-group ions density was significantly enhanced during the FTE shower intervals compared to the non-FTE shower intervals, and the enhancement was concentrated on the newly opened magnetic field lines, which formed a solar wind entry layer at the equatorward boundary of the northern cusp. The density of $Na^+$-group ions in the entry layer was approximately 0.6 $cm^{-3}$ or higher (Figure 4b), while the density was $\lesssim$0.3 $cm^{-3}$ in the similar northern cusp region during the non-FTE showers (Figure 4a). The magnetic field topology during the non-FTE shower was smooth without signatures of magnetic reconnection.

Figure 5 shows the distributions of $Na^+$-group ions during FTE shower intervals along with the magnetic local time (MLT) and magnetic latitude (MLAT) on the dayside magnetosphere. The figure shows that $Na^+$-group ions were generally found at an MLT of 09:00 to 15:00 and MLAT from 20° to 75°. The $Na^+$-group ions were concentrated in MLAT from 40° to 65° and MLT from 12:00 to 13:30 in the post-noon sector and were concentrated in MLAT from 40° to 50° and MLT from 9:30 to 12:00 in the pre-noon sector.

### 4.2. Solar Wind Entry Layer

The FTE shower opened a solar wind entry layer, in which the solar wind particles could be channeled into the magnetosphere and impacted the planet's surface at the cusp causing sputtering. Figure 6 traces the open magnetic field lines in the entry layer (see Appendix B on how to trace field lines) and shows the densities of $H^+$, $Na^+$-group ions, and $He^{++}$ as a function of the magnetic field intensities ($B_{mag}$). The densities of $H^+$ (around 29 $cm^{-3}$) and $He^{++}$ (around 0.6 $cm^{-3}$) at the "start" of the entry layer were high and decreased until the "shoulder" of the entry layer, where the density was about 1 $cm^{-3}$ for $H^+$ and about 0.01 $cm^{-3}$ for $He^{++}$. After the "shoulder", the densities of $H^+$ and $He^{++}$ increased almost linearly with the $B_{mag}$ till the "footprint" of the entry layer. The decrease of $H^+$ and $He^{++}$ density from the "start" to the "shoulder" implied that most of the solar wind ions transferred poleward, which would form the plasma mantle [*DiBraccio et al.*, 2015; *Jasinski et al.*, 2017; *Sun et al.*, 2020a]. The densities of $H^+$ and $He^{++}$ were linearly correlated with the $B_{mag}$ suggesting that these solar wind ions adiabatically moved along the flux tubes after passing the "shoulder", $n_{density} / B_{mag}$ = constant, i.e., the plasma content was conserved along the same flux tube, and the densities of $H^+$ and $He^{++}$ near the surface were approximately 5 $cm^{-3}$ and 0.15 $cm^{-3}$, respectively. As discussed above, since the energy of the protons was around 1 keV, it would take around 10 seconds for the protons with a pitch angle of 45° to travel from the subsolar magnetopause to the surface underneath the cusp (about 4000 km). Therefore, the solar wind entry layer was rapidly formed in less than a minute. The precipitation rates for $H^+$ and $He^{++}$ can be estimated to be $6.2 \times 10^{12}$ $m^{-2}$ $s^{-1}$ and $3.89 \times 10^{11}$ $m^{-2}$ $s^{-1}$. Meanwhile, the surface area of the cusp was determined to be approximately $2.65 \times 10^{12}$ $m^2$ (see Appendix C on how the surface area of the cusp is determined), the total precipitation rate was estimated to be $1.64 \times 10^{25}$ $s^{-1}$ for $H^+$, and is $1.0 \times 10^{24}$ $s^{-1}$ for $He^{++}$.

The precipitation rate of $H^+$ was around an order of magnitude higher than the average precipitation rate over the cusp obtained by *Winslow et al.* [2012], which confirmed that intense sputtering occurs during FTE shower intervals at Mercury. The precipitation rate of $He^{++}$ was around an order of magnitude lower than that of the $H^+$, but the $He^{++}$ ions could play a significant role in solar wind sputtering [*Szabo et al.*, 2018]. MESSENGER did not provide measurements of low-energy electrons, however, electrons should be precipitated simultaneously at a similar or even higher rate than the $H^+$.

### 4.3. Outflowing $Na^+$-group Ions

Figure 7 shows the $Na^+$-group ions pitch angle-energy distributions near the planet's surface beneath the northern cusp for the non-FTE shower intervals (upper panels) and FTE shower intervals (lower panels). The distribution in Figure 7 integrates over the FIPS measurements in the MLT from 09:00 to 15:00, MLAT from 55° to 70°, and altitude from 0 km to 244 km (0.1 $R_M$). In Figures 7a and 7c, most of the ions were concentrated between pitch angles from 60° to 150°. The distributions of scan numbers of FIPS were shown in Figures 7b and 7d, which show that FIPS generally covered all pitch angles relative to the magnetic field with the accumulation of the measurements from many MESSENGER's orbits.

The outflowing ions were measured with a range of energies (up to a few keV) but have the highest fluxes at relatively low energies (<400 eV). Table 1 shows the fluxes of $Na^+$-group ions measured at different pitch angles obtained from Figures 7a and 7c. We have integrated the fluxes for $Na^+$-group ions with pitch angles from 60° to 150° and obtained particle fluxes of $7.41 \times 10^{10}$ $m^{-2}$ $s^{-1}$ and $1.11 \times 10^{11}$ $m^{-2}$ $s^{-1}$ for non-FTE shower intervals and FTE shower intervals, respectively. The flux of $Na^+$-group ions was enhanced around 50% during FTE shower intervals.

## 5. Discussion

### 5.1. Solar Wind Sputtering Corresponds to Outflowing $Na^+$-group Ions

The enhanced $Na^+$-group ions over the northern cusp were outflowing from the planet's surface. These outflowing $Na^+$-group ions should be feed by ions released from the planet's surface. In this section, we discuss the processes that could release particles out of the planet's surface and try to find out which process generating the outflowing $Na^+$-group ions.

### 5.1.1. Thermal or Photon Stimulated Desorption, and Micrometeoroid Impact Vaporization

Thermal or photon stimulated desorption, and micrometeoroid impact vaporization cannot directly produce the enhanced $Na^+$-group ions in the entry layer of the cusp. Firstly, no evidence shows that they can specifically impact the cusp region, instead of thermal and photon stimulated desorption would generate an exospheric peak at the subsolar of the surface [*Killen et al.*, 2007; *Domingue et al.*, 2014]. Micrometeoroid impact vaporization would be higher near the apex of Mercury's orbit (on the dawnside hemisphere) and other sporadic micrometeoroid impact vaporizations are expected to randomly impact the planet's surface, which is unlikely distributed simultaneously as double peaks at northern and southern hemispheres [*Pokorný et al.*, 2018]. Secondly, the processes of thermal and photon stimulated desorptions correspond to long-term variations (days), i.e., their time scales are much longer than the one-minute response time during FTE showers. Thirdly, although micrometeoroid impact vaporization can cause a response on the timescale similar to the FTE showers [*Mangano et al.*, 2007], they should not be correlated with the magnetospheric activity that is caused by FTE showers.

Furthermore, the energy of the outflowing $Na^+$-group ions ranged from a few hundred eV to a few keV, which was much higher than the energy of neutral Na (up to a few eV) observed by MESSENGER's Mercury Atmospheric and Surface Composition Spectrometer (MASCS) [*McClintock & Lankton*, 2007]. These neutral Na exosphere is formed due to the thermal or photon stimulated desorption, and the temperature of neutral Na near the subsolar point is around 1200 K [*Cassidy et al.*, 2015] corresponding to an energy of around 1 eV, which is much lower than the observed outflowing $Na^+$-group ions. This is further evidence that the outflowing $Na^+$-group ions are not generated by the thermal or photon stimulated desorption.

### 5.1.2. Electron-Stimulated Desorption and Electron Impact Ionization

The solar wind contains electrons of high fluxes, which is higher than the values of proton fluxes. Electron-stimulated desorption is another important source for releasing Na neutral [*Yakshinskiy & Madey*, 2000; *McLain et al.*, 2011], and $Na^+$ and potassium ions ($K^+$) [*McLain et al.*, 2011]. Ions resulting from the electron-stimulated desorption are unlikely to have high velocities or large gyro-radii and would therefore remain attach to the planet's surface, which might enhance the effectiveness of the ion sputtering.

The electron impact ionization can contribute to the ionizations of neutrals in the atmosphere of Comets [*Cravens et al.*, 1987], Venus and Mars [*Ramstad & Barabash*, 2021]. In Figure 8, we make an estimation of electron ionization of neutral Na near Mercury's surface in the solar wind entry layer. The solar wind electron density is assumed to be comparable to the solar wind proton number density, $n_e \sim 30$ cm$^{-3}$, and the electron temperature is assumed to be 100 eV. The ionization frequency due to electrons can be calculated from [*Cravens et al.*, 1987]

$$R_{ioni} = n_e \int_{v_{pot}}^{\infty} v_e \sigma_s(v_e) f(v_e) 4\pi v_e^2 dv_e$$

$v_{pot}$ is the velocity corresponding to the ionization potential of Na (around 5 eV); $\sigma_s$ is the electron ionization cross-section of Na, which is a function of the electron velocity ($v_e$). Here we employ the $\sigma_s$ from [*Lotz*, 1967]. The ionization frequency is estimated to be $1.0 \times 10^{-6}$ s$^{-1}$. Considering a surface density of $1.0 \times 10^{4}$ cm$^{-3}$ and the scale height of 100 km for the neutral Na [*Cassidy et al.*, 2015] (Figure 8a), the production rate of $Na^+$ at the altitude of ~ 100 km is estimated to be ~ $2.0 \times 10^{-3}$ cm$^{-3}$ s$^{-1}$ (Figure 8b). The surface density of neutral Na in Figure 8a correspond to the upper value of neutral Na near the terminator of Mercury. The scale height corresponds to an average value [see, *Cassidy et al.*, 2015].

If the bombarding from the solar wind electrons lasts around 3 s, the ionized $Na^+$ can have a density of approximately 0.01 cm$^{-3}$. The cusp filament lasts around 3 s on average [*Poh et al.*, 2016], which corresponds to the bombarding associated with solar wind precipitation. The observed density of $Na^+$ was 0.6 cm$^{-3}$, which indicates that the density of $Na^+$ was around 2.09 cm$^{-3}$ (0.6 cm$^{-3}$ × $4\pi$ / $1.15\pi$). Therefore, we conclude that electron ionization contribute less than 1% of the $Na^+$ observed in the solar wind entry layer.

### 5.1.3. Solar Wind Ions Sputtering

In Figure 6, from the “start” to the “shoulder”, the densities of solar wind $H^+$ and $He^{++}$ decreased indicating that they convected poleward. However, the enhancement of the $Na^+$-group ions suggests the $Na^+$-group ions were piled up in the entry layer from the start to the shoulder, which indicated that the $Na^+$-group ions were continuously generated and outflowing from the planet’s surface. Those outflowing $Na^+$-group ions in Figure 7 were close to the planet’s surface with an altitude from 0 to 244 km. The sputtering of solar wind energetic ions can release atoms/ions from the planet’s regolith with energies up to a few hundreds eV [*Sigmund*, 1969; *Sieveka & Johnson*, 1984; *Hofer*, 1991; *Mura et al.*, 2007], which was consistent with the high energy of a few hundred eV to a few keV of the $Na^+$-group ions in Figure 7.

Another feature was that the flux of the $Na^+$-group ions was clearly enhanced (> 50%) during FTE shower intervals, which was another evidence that they were generated by the solar wind sputtering. The $Na^+$-group ions were less dense but did not completely disappear during the non-FTE shower intervals. These could be because i) the solar wind sputtering did not completely disappear as the cusp existed during both FTE shower intervals and non-FTE shower intervals; ii) those photoionized ions were accelerated by the solar wind and transferred into the magnetosphere [*Sarantos et al.*, 2009; *Raines et al.*, 2014; *Wurz et al.*, 2019; *Jasinski et al.*, 2021b].

Considering the precipitation rate of $H^+$ is $6.2 \times 10^{12}$ $m^{-2}$ $s^{-1}$ and the outflowing particle flux of the $Na^+$-group ions is $1.11 \times 10^{11}$ $m^{-2}$ $s^{-1}$ during FTE shower intervals, the yield for the sputtered $Na^+$-group ions can be calculated to be approximately 1.8%. Those sputtered ions would be firstly tied to the magnetic field line in the cusp by the Lorentz force and be energized by the magnetospheric convection electric field. A recent test particle simulation has shown that $Na^+$ test particles can gain up to hundreds of eV energy from the magnetospheric convection electric field [*Glass et al.*, 2021]. When the sputtered ions move anti-planetward along the magnetic field lines, i.e., outflowing, they would be subject to centrifugal acceleration [*Delcourt et al.*, 2012]. When the outflowing ions reach higher altitudes, they would encounter the solar wind convection electric field and be accelerated further (also see [*Raines et al.*, 2014; *Glass et al.*, 2021] and the high energy $Na^+$-group ions in Figure 1b).

## 5.2. Ion escaping channel of Mercury's exosphere

The integration over the entry layer with MLT from 06:00 to 18:00 gave a $Na^+$-group ions content of around $1.0 \times 10^{25}$. Considering a convection speed of 200 km/s and a scale of 1.36 $R_M$ of the entry layer, the average transport rate of $Na^+$-group ions from the dayside to the nightside can be estimated to be around $6.0 \times 10^{23}$ $s^{-1}$. Note that i) we used the Alfven speed at the start of the entry layer (approximately 230 km/s) to be the approximate convection speed; ii) this transport rate only considered the $Na^+$-group ions in the entry layer in the northern hemisphere. The whole transport rate considering both northern and southern hemispheres should be double the $6.0 \times 10^{23}$ $s^{-1}$ and was approximately $1.2 \times 10^{24}$ $s^{-1}$.

The transport rate of the $Na^+$-group ions ($1.2 \times 10^{24}$ $s^{-1}$) from the dayside to the nightside is comparable to the neutral Na escape rate (0.5 to $1.3 \times 10^{24}$ $s^{-1}$) in the Na tail [*Schmidt et al.*, 2010], which is primarily caused by the radiation-pressure-induced acceleration [*Ip*, 1986] and therefore is highly variable along Mercury's year. However, it is not clear what proportion of the estimated $Na^+$-group ions transport rate is due to $Na^+$. In the northern hemisphere of Mercury, the surface chemical composition (in wt%) for Na is 5.74%, Mg is 7.55%, Al is 6.04%, Si is 30.19% [*McCoy et al.*, 2018]. The neutral Na, therefore, forms around 11% of all four of these species at the surface. If we assumed that $Na^+$ ion was present in the same proportion as the surface composition of the $Na^+$-group species, then the transport rate of $Na^+$ ion was around $1.3 \times 10^{23}$ $s^-$

$^{1}$, which was several times lower than the escape rate of neutral Na in the Na tail. We note that there was no evidence that the $Na^+$ ion was in the same proportion as the neutral Na at the surface. A lack of *in situ*/laboratory experiments in this area meant we had to make this simple assumption.

The transport of $Na^+$-group ions during the FTE showers was likely an escape channel for Mercury's planetary atoms, which was driven by the solar wind-magnetosphere-surface coupling process and was different from the constant exospheric sodium loss due to the photoionization of the sodium exosphere. Photoionization removes approximately 0.9-4 × $10^{24}$ atoms/s of $Na^+$-group ions from the exosphere, with variations driven by seasons [*Jasinski et al.*, 2021b]. Our study focuses on short-term minutes timescales of the high latitude regions, while photoionization is long-term seasonal variations of the global exosphere. Ions escape is a common feature at the terrestrial planets. The $O^+$ ions escape from the dayside polar cap region at Earth at rates of $10^{24}$ to $10^{26}$ $s^{-1}$ [*Slapak et al.*, 2017; *Slapak et al.*, 2018]. The energetic ion plume of escaping $O^+$ ions observed in the induced magnetospheres at Mars is at the rate of $10^{24}$ to $10^{25}$ $s^{-1}$ [*Lundin et al.*, 2013] and Venus at approximately $10^{25}$ $s^{-1}$ [*McComas et al.*, 1986]. For those planets without a global intrinsic magnetic field, i.e., Mars and Venus, the escape ions are ionized by the solar ultraviolet (UV), electron impact ionization, or due to charge exchanges, which forms a constant escape channel [*Dubinin et al.*, 2011; *Ramstad & Barabash*, 2021]. The escape of $O^+$ at Earth depends on the solar wind parameters [*Slapak et al.*, 2017; *Schillings et al.*, 2019], which is similar to the escape channel of $Na^+$-group ions found in this study.

### 5.3. Influence on the neutral exosphere

In this section, we discuss how the solar wind sputtering influence the neutral Na exosphere at Mercury. At first, we estimate the surface density of neutral Na by considering the surface release flux of the $Na^+$-group ions.

If we assumed that the $Na^+$ formed 11% of the $Na^+$-group ions that were released from the surface by solar wind particles during FTE shower intervals (Figure 7c and Table 1), similar to the previous assumption, then the release flux of $Na^+$ ions in the northern hemisphere was around 1.22 × $10^{10}$ $m^{-2}$ $s^{-1}$. Since the sputtered $Na^+$ ions account for only 5% to 10% of the sputtered atoms [*Benninghoven*, 1975; *Hofer*, 1991], the sputtered neutral Na would approximately be an order of magnitude higher, which is around 2.4 × $10^{11}$ $m^{-2}$ $s^{-1}$. The exospheric density of the neutral Na at the surface ($n_{surf}$) can be estimated from

$$n_{surf} = f_{Na}/v_{release}$$

, where $v_{release}$ is the release velocity of neutral Na. The sputtering energy spectrum peaks at a few eV. Thus, we consider $v_{release}$ to be 3 km/s on average. As a result, $n_{surf} \sim 8.0 \times 10^7$ $m^{-3}$.

The exospheric surface densities of neutral Na range from $10^9$ to $10^{11}$ atoms/$m^3$ near Mercury's subsolar point and from $10^8$ to $10^{10}$ atoms/$m^3$ near the terminator [*Cassidy et al.*, 2015]. Hence, the FTE shower, on average, could likely enhance a considerable portion ($\gtrsim$ 10%) of the neutral Na in the cusp region through sputtering in minutes, which can likely cause the short-term variations of the Na emissions observed by ground-based telescope [*Massetti et al.*, 2017; *Orsini et al.*, 2018]. Our study provides clear evidence that dayside magnetopause reconnection, specifically FTE showers, injects solar wind ions into the cusps and enhance the $Na^+$-group ions in the high latitude magnetosphere. However, the causes of the short-term variability of the neutral Na exosphere could be more complex. This study provides a candidate, specifically FTE showers, for causing the short-term variations, which does not exclude that other processes might additionally cause short-term variations of the neutral Na exosphere.

Secondly, we can obtain the surface density of neutral Na by considering the estimated impact of solar wind proton fluxes and the known parameters derived for sputtering from the analytical model or laboratory. In the exospheric circulation models [*Mura et al.*, 2007; *Orsini et al.*, 2021], similar solar wind impact flux ($10^{13}$ $m^{-2}s^{-1}$) can produce an exospheric surface density of neutral Na of $10^8$ $m^{-3}$. Moreover, we can employ the sputtering yield derived from laboratory experiments. The estimated solar wind proton impact flux is $f_{impact} \sim 1.0 \times 10^{13}$ $m^{-2}$ $s^{-1}$. The sputtered neutral Na,

$$F_{Na} = f_{impact} \times y \times C$$

,where y is the yield (number of atoms released for each impacting ion) with a maximum measured value of 0.08 [*Johnson & Baragiola*, 1991; *Lammer et al.*, 2003] and C is the relative surface composition for Na, which at maximum is 0.06 [*Peplowski et al.*, 2014]. Hence, the maximum $F_{Na}$ is around $4.8 \times 10^{10}$ $m^{-2}$ $s^{-1}$. Then, dividing by the $v_{release}$, the $N_{SURF}$ is estimated to be approximately $1.6 \times 10^7$ $m^{-3}$.

There are differences between the models and the estimation based on the sputtered $Na^+$. The differences are not large and are within an order of magnitude. These differences could be due to several processes that have not been reproduced in laboratory studies. For example, i), the production rate of neutral Na depends on several factors, including temperature, the composition of the surface, and mineralogy [*Killen et al.*, 2007]. *Weider et al.* [2015] provided the global mapping of major elements on the surface of Mercury. However, the mineralogy about surface bounds still has not been well derived; ii) the solar wind includes alpha ions ($He^{++}$) with a precipitation rate of around $1.0 \times 10^{24}$ $s^{-1}$. The $He^{++}$ ions could enhance the yield of the sputtering [*Szabo et al.*, 2018]. The solar wind also includes a large portion of electrons. The electron stimulated desorption could also affect the release of neutrals and ions from the planet's surface.

## 6. Further Impact and Future Mission

The results from this study can influence a variety of aspects. Not only is solar wind-magnetopause reconnection important for directly influencing the exospheric dynamics and planetary ion escape at Mercury, but also magnetic reconnection can input explosive energy from the solar or a stellar wind into the magnetosphere of planets or an exoplanet under intense external driving [*Barclay et al.*, 2013] similar to Mercury. For example, Ganymede, one of the Galilean moons, has a global magnetic field [*Kivelson et al.*, 1996] and is located in a sub-Alfvénic corotation flow in Jupiter's magnetosphere. The sub-Alfvénic flow refers to flow speed smaller than the background Alfvén speed, and therefore corresponds to low $M_A$. In a recent simulation study, *Zhou et al.* [2019]; [and *Zhou et al.*, 2020] show that magnetic reconnection can frequently generate magnetic flux ropes on the magnetopause and input a significant amount of energy into Ganymede's magnetosphere. Exoplanets with a global magnetic field close to their primary stars could be exposed to similar low $M_A$ stellar wind [*Ip et al.*, 2004]. At those planets, intense magnetic reconnection can be expected to occur that leads to efficient transport of plasma and energy from the stellar wind into the planet's atmosphere or surface, which can facilitate atmospheric escape, as simulated by *Egan et al.* [2019], and therefore affect the habitability of planets and exoplanets.

A joint ESA-JAXA mission, BepiColombo [*Milillo et al.*, 2020], consisting of the Mercury Planetary Orbiter (MPO) and the Mercury Magnetospheric Orbiter (Mio), made its first flyby of Mercury in October 2021 with Mercury orbit insertion scheduled in late 2025 or early 2026. BepiColombo will provide many comprehensive measurements on Mercury's magnetosphere and exosphere, especially those higher resolutions measurements for different ion species, i.e., Mercury Plasma Particle Experiment (MPPE) [*Saito et al.*, 2021] onboard Mio, and neutrals, i.e., Search for Exospheric Refilling and Emitted Natural Abundances (SERENA) [*Orsini et al.*, 2021]. Moreover, MPO and Mio will have much broader altitudinal coverage of both northern and southern cusps than MESSENGER was able to achieve. At times, one spacecraft will serve as a solar wind monitor to the other spacecraft inside the magnetosphere. The impact of magnetopause reconnection on Mercury's exospheric dynamics will be investigated in much detail.

# Appendices

## Appendix A. Hall magnetohydrodynamics (MHD) Simulation

The simulation result is shown in Figure 1 (right panel) and was extracted from a global Hall-MHD simulation of Mercury's magnetosphere. The simulation was performed using the Hall-MHD version of the BATSRUS code [*Tóth et al.*, 2012] that enables us to properly simulate fast magnetic reconnection on the magnetopause. In our resistive, Hall-MHD treatment, the generalized Ohm's law reads as: $\vec{E} = \eta\vec{J} - \vec{u} \times \vec{B} + \frac{1}{ne}\vec{J} \times \vec{B}$, where $\eta\vec{J}$ is the resistivity term, $\vec{u}$, $n$ and $\vec{B}$ are plasma velocity, density, and magnetic field, respectively. The Hall-term (the last term on the right-hand side of the above equation) is important for plasma dynamics on scales shorter than ion inertial scale lengths, but greater than electron inertial scale lengths. The magnetic field lines are frozen to the electron fluid but not to the ion fluid due to the Hall effect. It has been demonstrated that Hall-MHD appears to be the minimal modification required for an MHD code to reproduce the fast reconnection process seen in particle and hybrid simulations [*Birn et al.*, 2001; *Chen et al.*, 2019]. Our global Hall-MHD model also electromagnetically couples Mercury's interior to the surrounding magnetosphere, allowing us to directly simulate the induction effect arising from Mercury's large-size conducting core [*Jia et al.*, 2015; *Jia et al.*, 2019]. This is achieved primarily through the resistivity term included in the generalized Ohm's law, for which different resistivity (or inversely conductivity) values are prescribed to represent different electrical properties of the planet's mantle and core.

## Appendix B. The derivation and the trace of the magnetic field lines

The measurements of the magnetic field vector during the FTE shower intervals and the non-FTE shower intervals are averaged over the dayside magnetosphere, respectively. In Figure 4 and Figure 6 the magnetic field lines are derived from the averaged magnetic field vectors within the local times from 10 MLT to 14 MLT. Figure B1 shows the spatial distributions of magnetic field intensity and the magnetic field lines during intervals of FTE showers and intervals of non-FTE showers. FTE shower intervals contain 1953 dayside magnetosphere crossings, and non-FTE shower intervals contain 1795 dayside magnetosphere crossings.

In Figure 6, we have traced the magnetic field lines in the solar wind entry layer, which is the shaded region between the closed and open magnetic field lines as shown on the left panel. The starting point is indicated by the "Start", and the endpoint the "Footprint". The magnetic field intensity and the densities of ion species are obtained through average over the entire entry layer within the MLT from 12 MLT to 14 MLT.

### Appendix C. The determination of the surface area of the cusp

The area of the northern cusp is determined from the distribution of the alpha ions ($He^{++}$) measured by FIPS. Figure C1 includes a spatial distribution of the alpha ions during the intervals of FTE showers and non-FTE-showers, respectively, along Mercury's magnetic local time (MLT) and magnetic latitude (MLAT). The northern cusp is defined to be the area in the high latitude with the densities of the alpha ions being larger than 0.03 $cm^{-3}$. The area of the cusp is calculated from $r \times \cos(\Delta MLONG) \times \Delta MLAT$, where $r$ represents the radial length, ΔMLONG is the width of the angle along the magnetic longitude (i.e., the MLT), and ΔMLAT is the width of the angle along the magnetic latitude. The ΔMLAT is not a constant along different MLT. We calculate the value of the area in each grid of MLT and then integrate them. The surface area (A) is estimated to be $3 \times 10^{12}$ $m^2$**.**

## Acknowledgements

The MESSENGER project was supported by the NASA Discovery Program under Contracts NASW-00002 to the Carnegie Institution of Washington and NAS5-97271 to The Johns Hopkins University Applied Physics Laboratory. We thank S. C. Solomon and the MESSENGER team for the successful operation of the spacecraft, and the instrument teams of FIPS from the University of Michigan, Ann Arbor, and MAG from the NASA Goddard Space Flight Center, Greenbelt, and the Johns Hopkins University Applied Physics Laboratory. We thank the team of SWMF/BATSRUS at the University of Michigan, Ann Arbor. MESSENGER data are available through the Planetary Plasma Interactions (PPI) Node of the NASA Planetary Data System (PDS) at https://pds-ppi.igpp.ucla.edu/. The magnetic field data measured by MAG is available at https://pds-ppi.igpp.ucla.edu/search/view/?f=yes&id=pds://PPI/MESS-E_V_H_SW-MAG-3-CDR-CALIBRATED-V1.0 , the ion data measured by FIPS is available at https://pds-ppi.igpp.ucla.edu/search/view/?f=yes&id=pds://PPI/MESS-E_V_H_SW-EPPS-3-FIPS-DDR-V2.0. The list of Mercury's dayside magnetopause crossings (3,748 crossings) made by MESSENGER from 11 March 2011 to 30 April 2015 is available at the supporting information of https://doi.org/10.1029/2020GL089784. The model data were obtained from simulations using the SWMF/BATSRUS code developed at the University of Michigan, which is publicly available at http://csem.engin.umich.edu/tools/swmf/. W. J. S. and J. A. S. were supported by NASA Grants 80NSSC21K0052, and 80NSSC18K1137. J. M. J. acknowledges support from the Jet Propulsion Laboratory, California Institute of Technology, under a contract with NASA as well as NASA's Discovery Data Analysis Program (grant number 80NM0018F0612). W. J. S. thanks Markus Fränz (MPI fuer Sonnensystemforschung) and Yong Wei (Institute of Geology and Geophysics, Chinese Academy of Sciences) for the helpful discussions.

## Author Contributions

W. J. S. and J. A. S. designed the work. J. A. S provided guidance. W. J. S. led the work, identified the events, conducted the data analysis of the dataset, and wrote the manuscript. R. M. D. provided help in the analysis of Figure 2, Figure 7, and Table 1. A. M. and S. O. contributed to the determination of neutral densities. X. J. and C. L. provided the Hall-MHD simulation of FTEs on Mercury's magnetopause. J. M. R., J. M. J. and R. M. D. provided knowledge of the FIPS instrument and the analysis of the ion measurements. All authors discussed and contributed to the manuscript.

## Figures

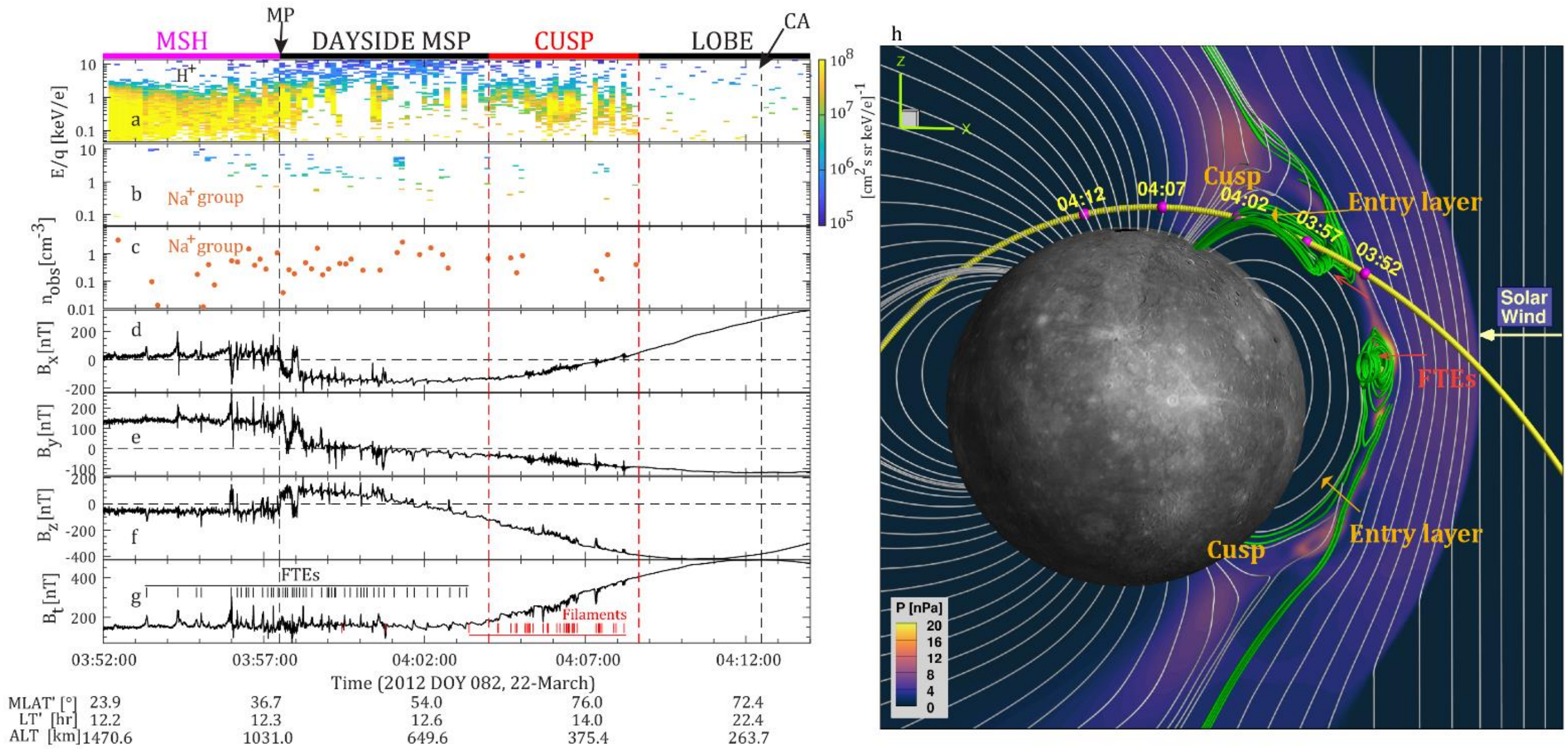

**Figure 1. Overview of the flux transfer event (FTE) shower observed by MESSENGER on 22 March 2012 (left) and the Hall-MHD simulation (right).** Left: (a), differential particle flux for protons (see Methods, Satellite and Instrumentation), (b) differential particle flux for sodium ion ($Na^+$) group (m/q = 21 to 30), (c) observed densities of $Na^+$-group. (d-f) three magnetic field components, $B_x$,, $B_y$, $B_z$, respectively, (g) magnetic field intensity $B_t$. The black and red ticks mark flux transfer events (FTEs) and cusp filaments, respectively. Observations in the magnetosheath (MSH), dayside magnetosphere (Dayside MSP), cusp and lobe are labeled at the top of the figure. Magnetic Latitude (MLAT), Local Time (LT), and Altitude (ALT) of the satellite are labeled underneath the figure. The dayside magnetopause (MP) crossing occurred at an altitude of ~ 990 km, close to the equator (~ 37.2°) and at local noon (~ 12:20 local time). Closest approach (CA) occurred at an altitude of ~ 263 km at high latitudes (~ 71.0°) and at local midnight (~ 22:35 local time). Right: (h) the BATSRUS Hall-Magnetohydrodynamics (MHD) simulation of Mercury's magnetosphere under similar interplanetary magnetic field conditions to the MESSENGER observations. In this snapshot, two FTEs with a clear helical magnetic field topology (green lines) appeared simultaneously at the dayside magnetopause. The solar wind entry layers were identified equatorward of the northern and southern cusps, respectively. The cusps are identified as the regions with higher thermal pressure. The yellow line represents MESSENGER's trajectory, which contains some time marks consistent with the left panel.

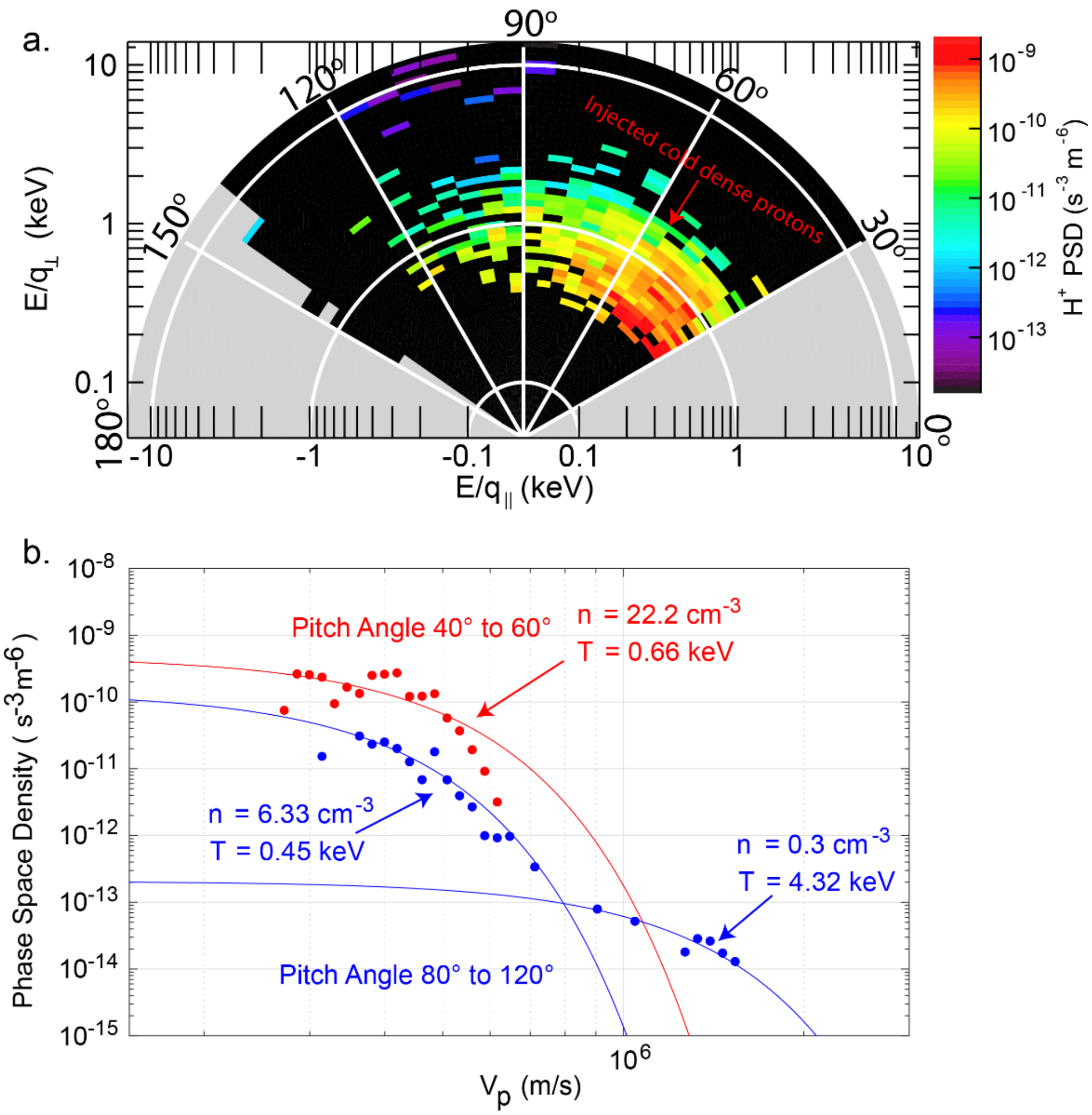

**Figure 2. FIPS measurements of the pitch angle-energy distribution inside the FTEs in the dayside magnetosphere during the MESSENGER event shown in Figure 1.** (a) The distribution of proton phase space density (PSD) versus pitch angle. This distribution integrates over all the unwrapped measurements inside the FTEs which the spacecraft crossed inside the magnetosphere. Pitch angle bin size is 5°. Grey regions indicate the pitch angles are out of the field-of-view (FOV) of FIPS in the period. (b) Gaussian fits on the protons in the perpendicular direction (pitch angles from 80° to 120°, in blue) and protons in the parallel direction (pitch angles from 40° to 60°, in red). Parallel direction contains a cold and dense population. Perpendicular direction contains two populations, one is cold and dense, which is identified to have been injected from the solar wind, and the other is a hot and tenuous magnetospheric population.

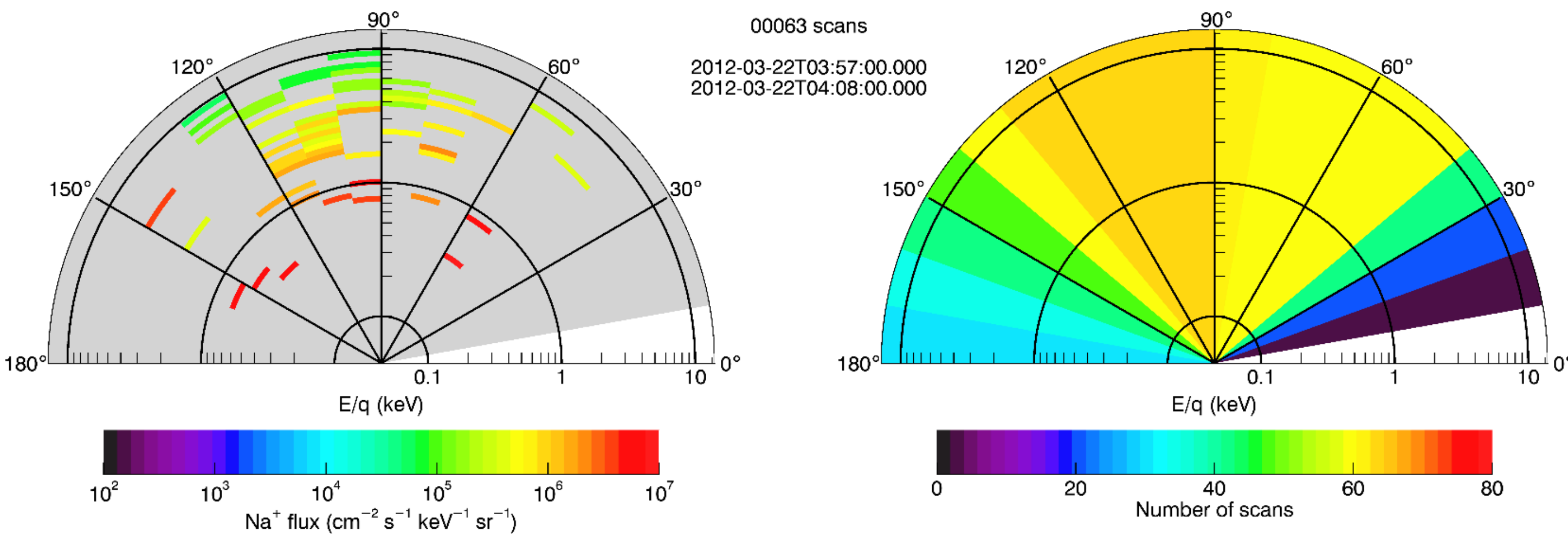

**Figure 3. FIPS measurements of the distribution of the $Na^+$-group ions during this period from 03:57 to 04:08, 22 March 2021 UTC**, corresponding to the MESSENGER event in Figure 1. (a) the distribution of the differential particle flux of $Na^+$-group ions versus pitch angles. (b) the distribution of the number of FIPS scans versus pitch angles. Pitch angle bin size is 5°.

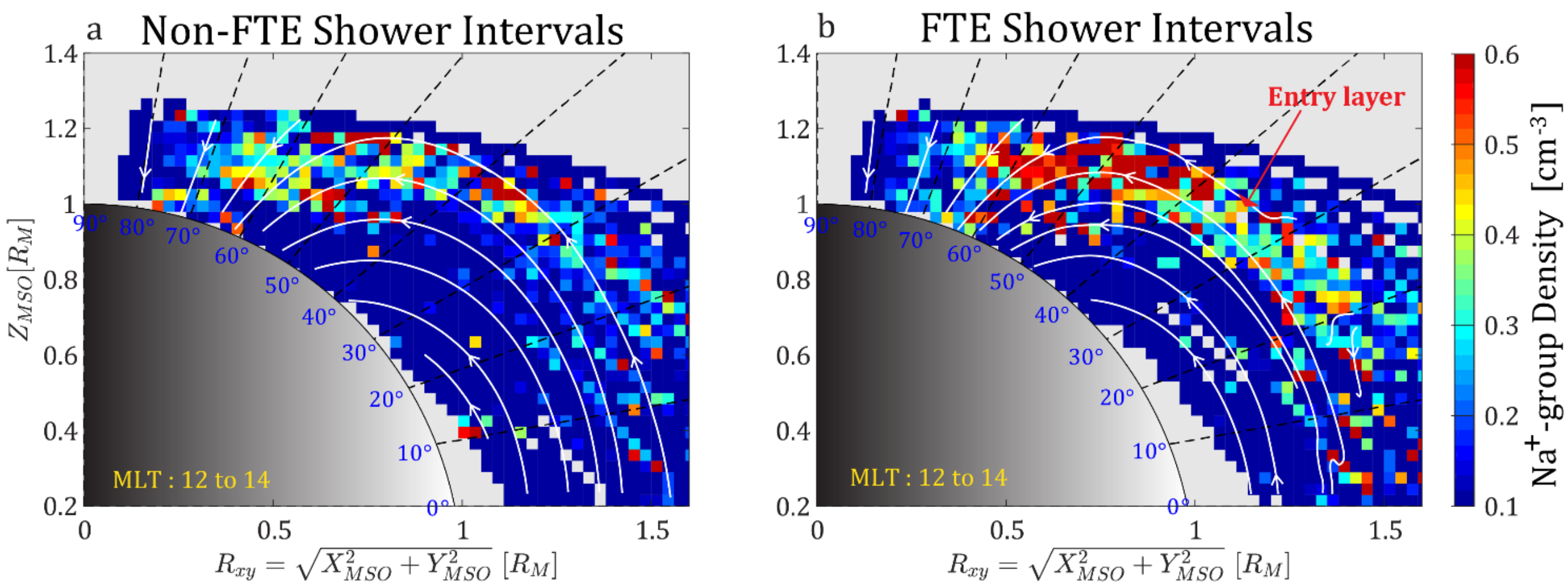

**Figure 4. The magnetic field topology as well as MESSENGER spatial distribution measurements of the sodium-group ($Na^+$-group) ions during (a) intervals without FTE showers and (b) intervals with FTE showers, shown in the $R_{xy}$-Z plane ($R_{xy} = \sqrt{X^2_{aMSO} + Y^2_{aMSO}}$).** Colors indicate the observed density of the $Na^+$-group. The white lines represent the magnetic field lines obtained through the average magnetic fields measured by MESSENGER during the intervals without FTE showers and with FTE showers, respectively (see Appendix B on how the field lines are derived). The solar wind entry layer (indicated by the red arrow in b) is determined from the magnetic field topologies during the intervals of FTE showers. In this figure, the measurements of the $Na^+$-group were limited in the magnetic local time (MLT) from 12:00 to 14:00.

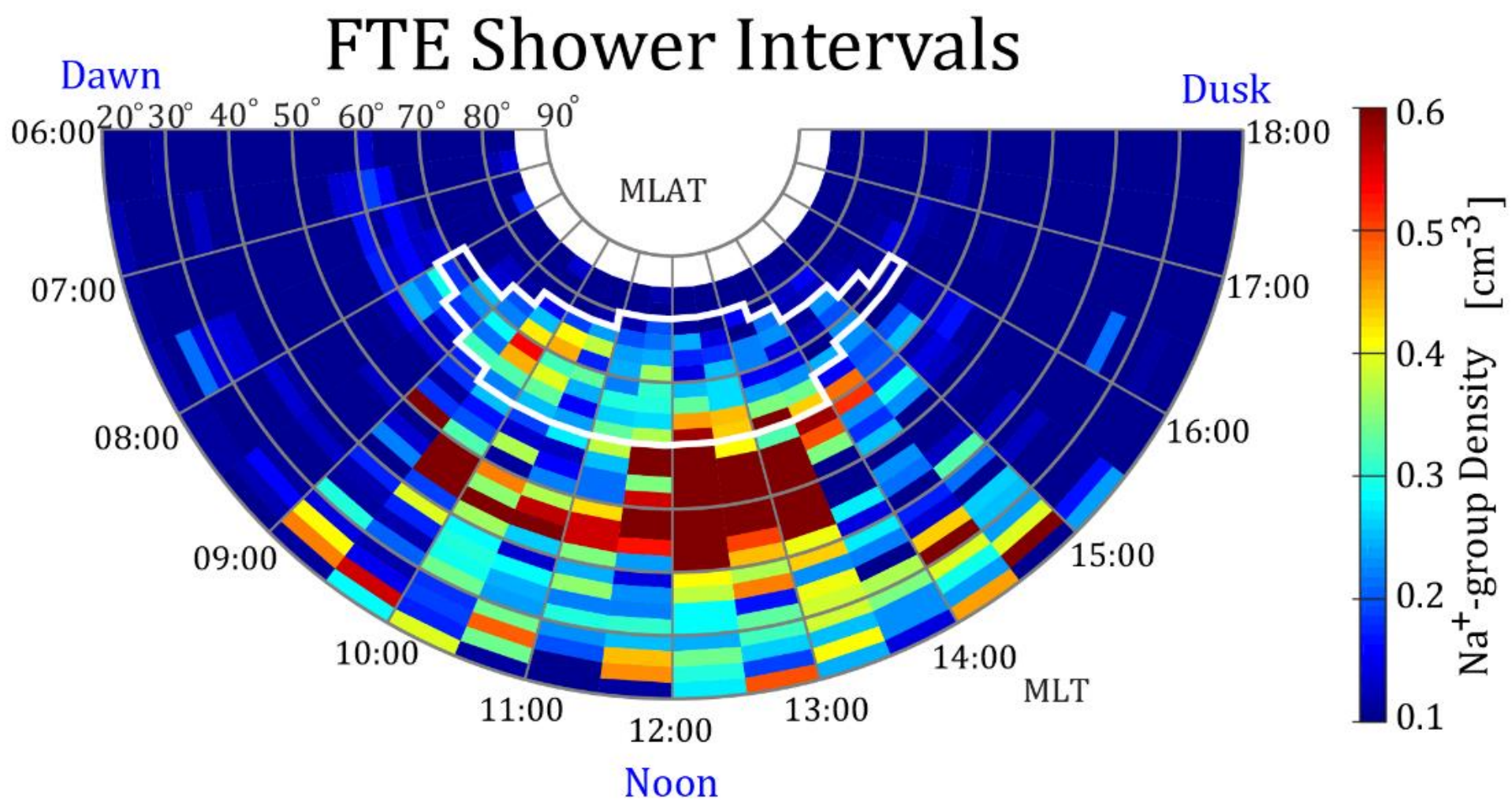

**Figure 5. Spatial distributions of sodium-group ($Na^+$-group) ions during intervals of FTE showers in the entry layer along with Mercury's magnetic local time (MLT) and magnetic latitude (MLAT).** Colors indicate the observed density of the $Na^+$-group ions. The intervals of FTE showers contain 1953 magnetopause crossings. The white contour includes the cusp which is determined from the spatial distributions of alpha ions ($He^{++}$, see Figure C1).

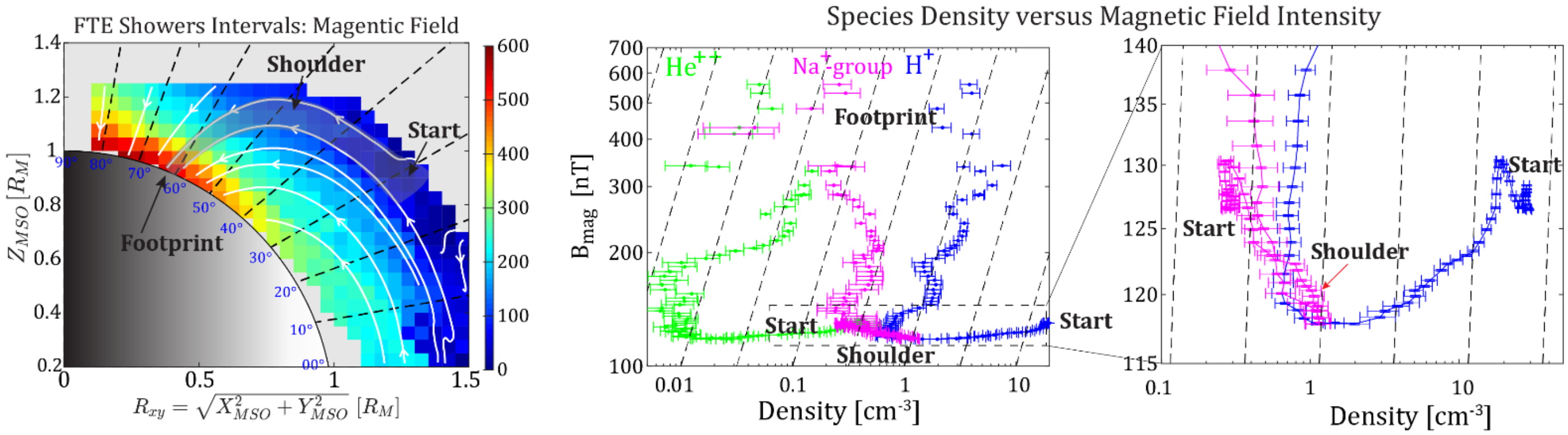

**Figure 6. The densities of sodium-group ($Na^+$-group) ions, protons ($H^+$), and alpha ions ($He^{++}$) in the solar wind entry layer on the equatorward boundary of the northern cusp during the intervals of FTE showers.** Left: the distribution of the averaged magnetic field intensity and the magnetic field lines during FTE shower intervals (the magnetic field lines are the same as shown in Figure 3b). The "start", "shoulder" and "footprint" of the entry layer are marked. Middle and right: the variations of the observed densities of $Na^+$-group (in magenta), $H^+$ (in blue), and $He^{++}$ (in green) in the entry layer along with the magnetic field intensity ($B_{mag}$). The magnetic field density is the averaged magnetic field intensity measured by MESSENGER in the entry layer. The dashed lines represent the linear correlation between $n_{density}$ and $B_{mag}$, i.e., $n_{density}$ / $B_{mag}$ = constant. The linear correlation suggests adiabatic travel of ions. In this figure, the measurements of the $Na^+$-group ions, $H^+$, and $He^{++}$ were limited in the magnetic local time (MLT) from 10:00 to 14:00.

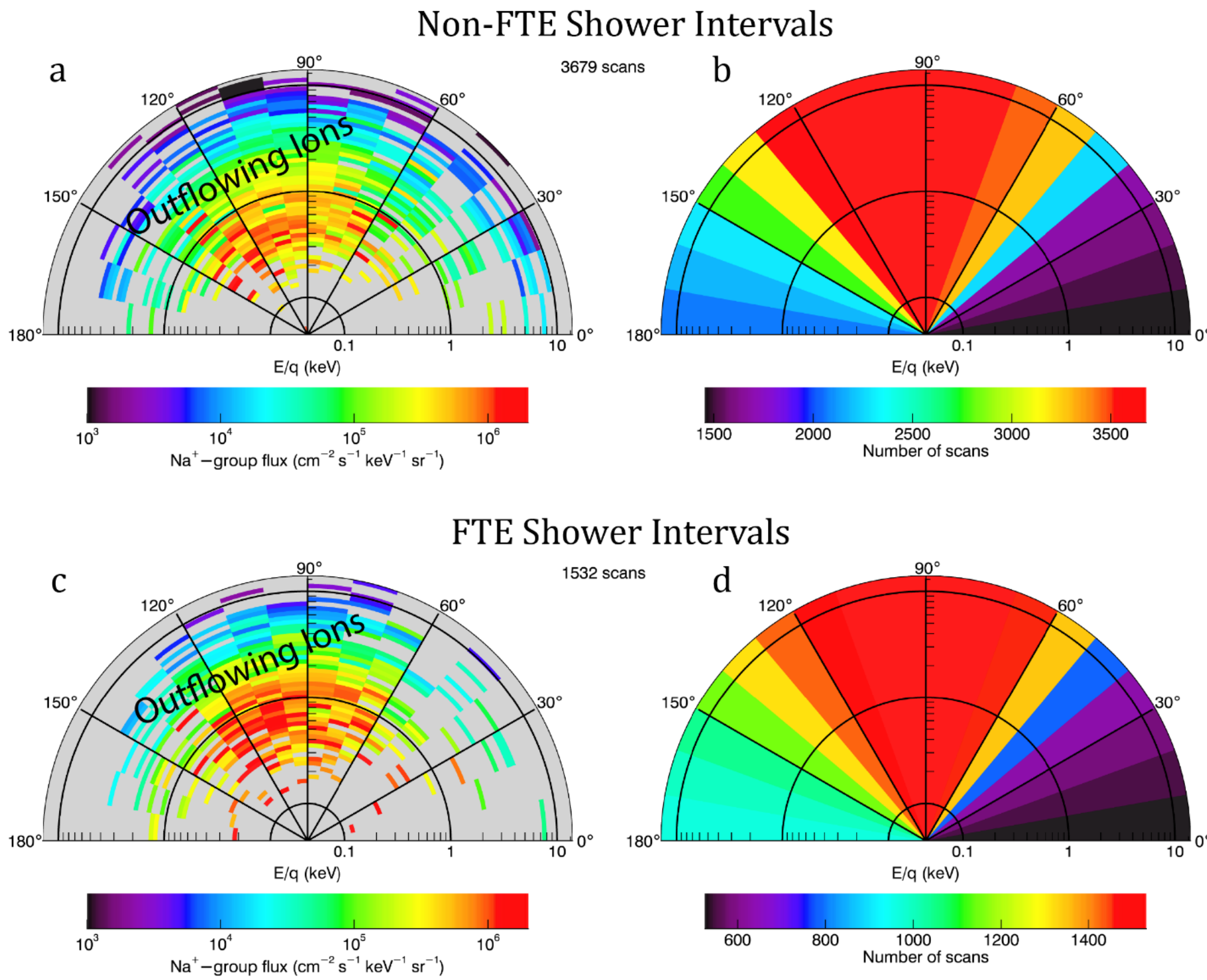

**Figure 7. FIPS measurements of the pitch angle-energy distribution of the Na$^+$-group near the planet's surface beneath the northern cusp.** The integration areas include MLT from 09:00 to 15:00, MLAT from 55° to 70°, and altitude from 0 km to 244 km (0.1 $R_M$). The pitch angle bin size is 10°. Upper panels (a and b) are for the non-FTE shower intervals and lower panels (c and d) are for FTE shower intervals. Panels on the left (a and c) show the distributions of the differential particle flux (cm$^{-2}$ s$^{-1}$ keV$^{-1}$ sr$^{-1}$). Panels on the right (b and d) show the distributions of the number of scans made by FIPS in each pitch angle bin.

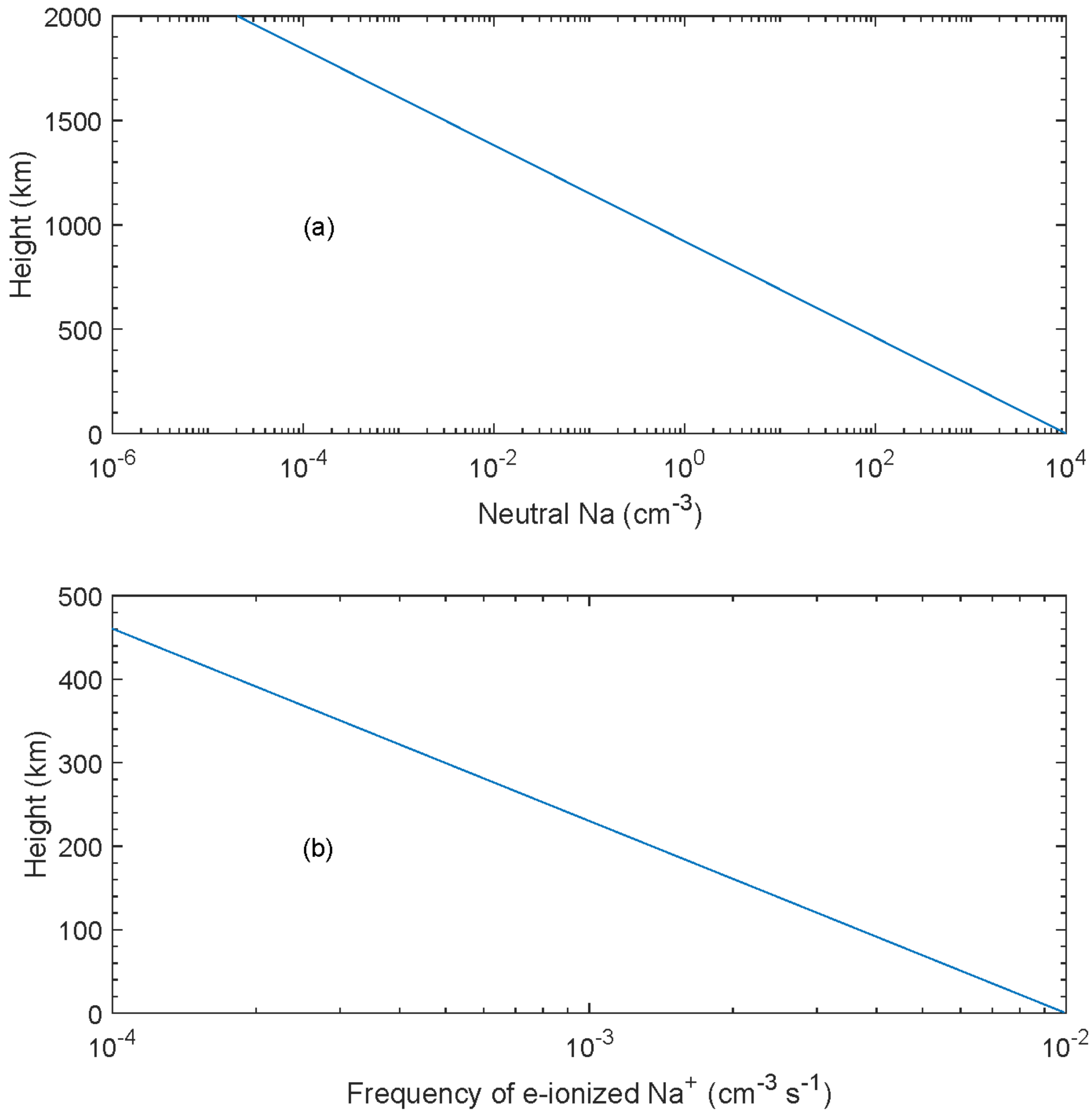

**Figure 8. An estimation of electron ionization of neutral Na.** Upper panel: the density profile of Na along with height, i.e., altitude; Bottom panel: the production rate of $Na^+$ due to electron ionization along with height. The surface density of the neutral Na is $1 \times 10^4$ $cm^{-3}$ with a scale height of 100 km. The electron temperature is set to be 100 eV. The neutral Na with the density of $1 \times 10^4$ $cm^{-3}$ corresponds to the upper value of neutral Na near the terminator of Mercury.

***Table 1. The flux of Na$^+$-group near the planet's surface underneath the cusp, which corresponds to the pitch angle-energy distribution in Figure 7.***

| Pitch Angle (°) | FTE shower intervals | | Non-FTE shower intervals | |
|---|---|---|---|---|
| | Flux ($cm^{-2}$ $s^{-1}$ $sr^{-1}$) | Number of Counts | Flux ($cm^{-2}$ $s^{-1}$ $sr^{-1}$) | Number of Counts |
| 0 to 10 | $3.89 \times 10^4$ | 1 | $1.31 \times 10^5$ | 4 |
| 10 to 20 | $9.08 \times 10^4$ | 2 | $7.02 \times 10^4$ | 9 |
| 20 to 30 | $2.49 \times 10^5$ | 60 | $1.74 \times 10^5$ | 25 |
| 30 to 40 | $1.70 \times 10^5$ | 5 | $1.43 \times 10^5$ | 18 |
| 40 to 50 | $1.51 \times 10^5$ | 10 | $1.99 \times 10^5$ | 26 |
| 50 to 60 | $3.29 \times 10^5$ | 15 | $6.0 \times 10^5$ | 34 |
| 60 to 70 | $1.06 \times 10^6$ | 48 | $6.33 \times 10^5$ | 74 |
| 70 to 80 | $1.22 \times 10^6$ | 100 | $8.28 \times 10^5$ | 159 |
| 80 to 90 | $1.43 \times 10^6$ | 137 | $8.76 \times 10^5$ | 197 |
| 90 to 100 | $1.52 \times 10^6$ | 176 | $9.46 \times 10^5$ | 242 |
| 100 to 110 | $1.68 \times 10^6$ | 179 | $9.89 \times 10^5$ | 250 |
| 110 to 120 | $1.63 \times 10^6$ | 138 | $9.40 \times 10^5$ | 186 |
| 120 to 130 | $1.17 \times 10^6$ | 77 | $9.02 \times 10^5$ | 128 |
| 130 to 140 | $1.09 \times 10^6$ | 33 | $8.59 \times 10^5$ | 62 |
| 140 to 150 | $2.87 \times 10^5$ | 21 | $6.34 \times 10^5$ | 37 |
| 150 to 160 | $1.80 \times 10^5$ | 14 | $9.64 \times 10^4$ | 18 |
| 160 to 170 | $8.11 \times 10^4$ | 7 | $6.53 \times 10^4$ | 11 |
| 170 to 180 | $1.09 \times 10^5$ | 3 | $2.43 \times 10^4$ | 2 |

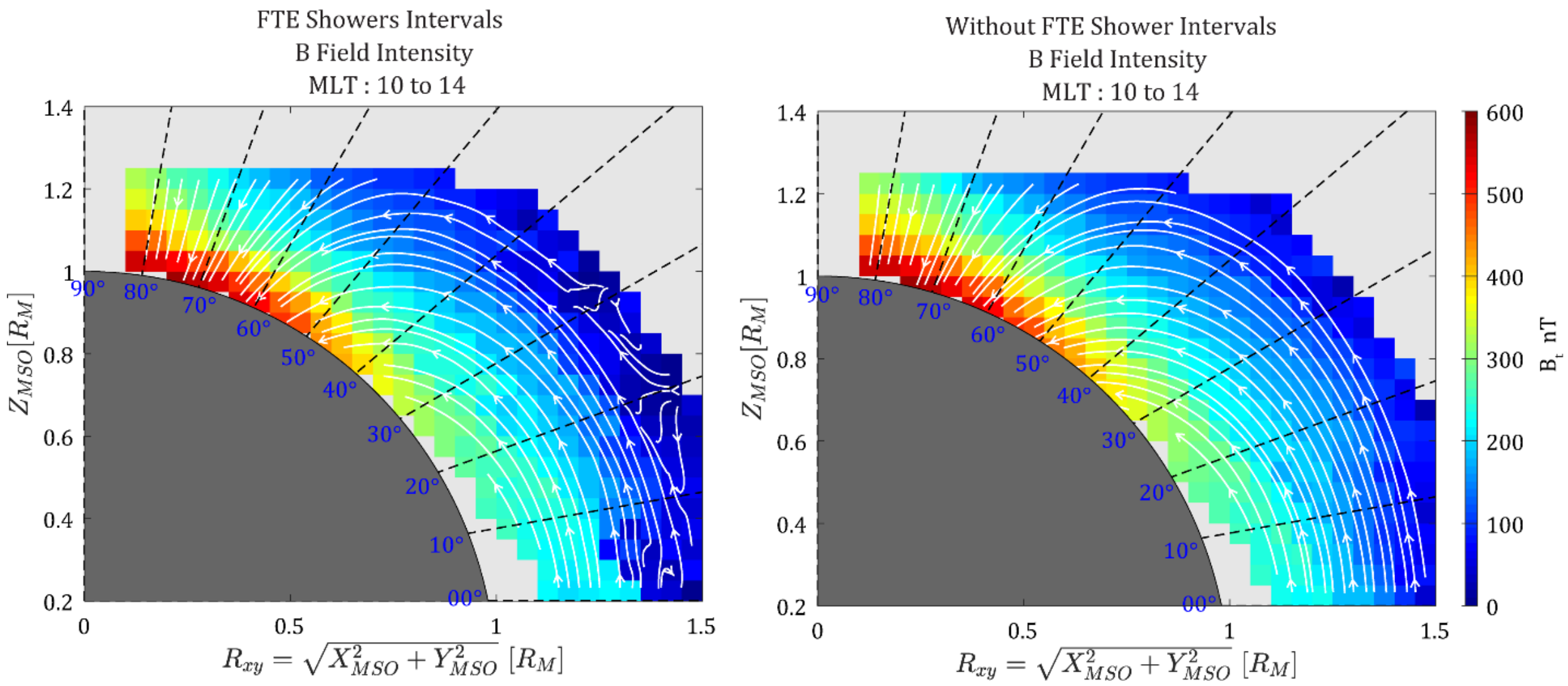

**Figure B1. Spatial distributions of magnetic field intensity and the derived magnetic field lines during intervals of FTE showers (left) and intervals of non-FTE showers (right) in the $R_{xy}$-Z plane ($R_{xy} = \sqrt{X_{aMSO}^2 + Y_{aMSO}^2}$).** Colors indicate the observed magnetic field intensity from MESSENGER, which were limited in the magnetic local time (MLT) from 10:00 to 14:00. The white lines represent the magnetic field lines derived through the averaged magnetic field intensities. The left figure during the intervals of FTE showers, there are few curved magnetic field lines on the outer edge, which should be signatures of magnetic reconnection or FTEs. On the right figure, the magnetic field lines are smooth.

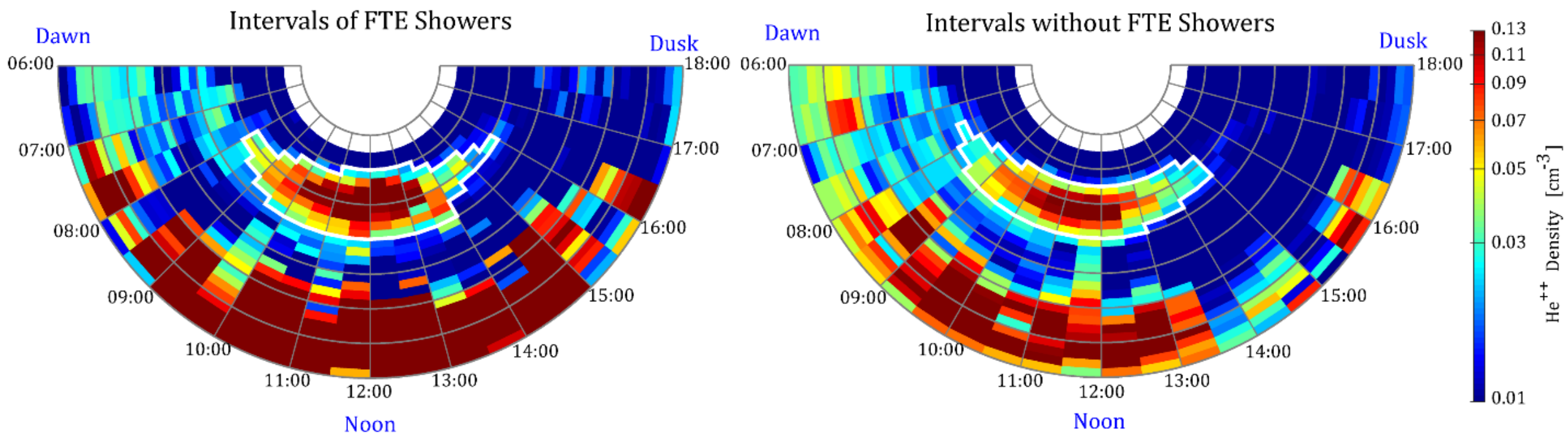

**Figure C1. Spatial distributions of Alpha ($He^{++}$) particles during the intervals of FTE showers (left) versus intervals without FTE showers (right) along Mercury's magnetic local time (MLT) and magnetic latitude (MLAT).** Colors indicate the observed density of $He^{++}$. The white contours include regions with $He^{++}$ density higher than 0.03 $cm^{-3}$.